\documentclass[twocolumn,aps,prb,showpacs,floatfix,
]{revtex4-1}

\usepackage{amsmath}
\usepackage{graphicx}
\usepackage{color}

\def\beq{\begin{eqnarray}}
\def\eeq{\end{eqnarray}}

\begin{document}

\title{Bound states in open coupled asymmetrical waveguides and quantum wires}

\author{Paolo Amore}
\affiliation{Facultad de Ciencias, CUICBAS, Universidad de Colima, Bernal
D\'{\i}az del Castillo 340, Colima, Colima, Mexico}

\author{Martin Rodriguez}
\affiliation{Facultad de Ciencias, CUICBAS, Universidad de Colima, Bernal
D\'{\i}az del Castillo 340, Colima, Colima, Mexico}

\author{C\'esar A. Terrero-Escalante}
\affiliation{Facultad de Ciencias, CUICBAS, Universidad de Colima, Bernal
D\'{\i}az del Castillo 340, Colima, Colima, Mexico}

\begin{abstract}
The behavior of bound states in asymmetric cross, T and L shaped configurations is considered.
Because of the symmetries of the wavefunctions, the analysis can be reduced to the case of an electron 
localized at the intersection of two orthogonal crossed wires of different width.
Numerical calculations show that the fundamental mode of this system remains bound for the widths that we have been able to study
directly; moreover, the extrapolation of the results obtained for finite widths suggests that this state remains bound even when the width 
of one arm becomes infinitesimal. We provide a qualitative argument which explains this behavior and that can be generalized to the lowest energy
states in each symmetry class. In the case of odd-odd states of the cross we find that the lowest mode is bounded when the width of the two arms
is the same and stays bound up to a critical value of the ratio between the widths; in the case of the even-odd states we find that the lowest 
mode is unbound up to a critical value of the ratio between the widths. Our qualitative arguments suggest that the bound state survives as 
the width of the vertical arm becomes infinitesimal.
\end{abstract}


\maketitle


\section{INTRODUCTION}

\label{sec:intro}

Trapped waves in open geometries have been known for sixty years now \cite{Ursell:1951}. 
Ursell \cite{Ursell:1952}, in a theoretical and experimental study of ``beach" waves in a particular semi--infinite canal, 
found that the system has discrete, continuous and mixed spectra, and predicted the existence of a confined 
resonance at a discrete frequency, while at a cutoff frequency the resonance extends a long way down the canal. 
These waves can be described by the Helmholtz equation subject to Dirichlet boundary conditions and
therefore similar confinement properties are expected in the solutions of analogous acoustic or electromagnetic 
problems. Indeed, about fifteen years after Ursell, 
long-lived resonance modes in open laser systems were reported\cite{Weinstein:1966}.
Even more, since this setup is equivalent to that of electrons in open configurations described by the time-independent Schr\"odinger equation, 
by 1984, several studies already suggested that confinement is also a feature of the two-dimensional transport of charge carriers 
in ultrafine metal and semiconductor devices\cite{Sakaki:1984}.
From then on, many theoretical and experimental studies have shown that, generally, by bending and crossing waveguides and quantum wires, bound (confined, trapped) 
states can be obtained with energies below the continuous spectrum\cite{Carini:1997}. 
Moreover, many similar quantum wire systems have been found with bound states embedded into the continuum\cite{Londergan:1999}. 

Bound states just below the continuum may strongly influence the scattering of waves and charge carriers. 
This allows to detect them and it is also the source of new phenomena and applications.
For instance, mesoscopic systems with similar geometries are common in nanoelectronics and it has been shown experimentally that they exhibit 
suppression (‘‘quenching’’) of the Hall resistance\cite{Roukes:1988}, and enhanced bend resistance\cite{Timp:1988}.
On the other hand, if the above mentioned bound states exist only due to a particular symmetry, when this symmetry is broken, resonances could be expected to arise.  
Therefore, the confinement of electromagnetic radiation at the intersection of coupled waveguides opens a window for a new kind of resonator,
which combines a very simplified spectrum to high performances\cite{Annino:2006}.
Similar phenomena should be also relevant for the performance of novel kinds of (cold) atom\cite{Bromley:2003}, phonon\cite{Qu:2004} and laser waveguides\cite{Hill:2007}.
Even more, the possibility of geometrically inducing confined states allows for the design of unused ways 
to create a Bose-Einstein condensate in a quasi-one dimensional Bose gas\cite{Exner:2005}.
Last but not least, besides the possible practical applications, 
it should be noted that such systems constitute a very interesting theoretical framework for studying the correspondence between classical and quantum dynamics.
Note that, in this kind of arrangements, classical motion of point particles with finite energy is typically unbound because there are not forbidden regions along the wires.
Therefore, a phase-space semiclassical approximation to dynamics seems to be not valid here, 
and this is particularly relevant if the corresponding classical motion exhibits chaos\cite{Exner:1996}.

A relevant question, from both the theoretical and experimental point of view, is how strongly depend the properties of the bound states 
on the perturbation of the structure and, this way, on the mechanical imperfections. 
Amongst other factors, is perhaps the symmetry of the configuration what plays the fundamental role since it controls the coupling with the propagating modes.
Some studies suggested that the isolated bounded modes are not affected by the level of symmetry and thus, 
they can exist under very general conditions\cite{Annino:2006b}.
Nevertheless, a comprehensive analysis of the role of symmetry is still lacking.

In a classical paper written more than twenty years ago Schult, Ravenhall and Wyld\cite{Schult89} 
studied the problem of an electron which can freely move in a two dimensional symmetric region of orthogonally crossed 
wires of finite width and infinite length.
They showed that the ground state of this system corresponds to a state where the electron is confined 
in the region of the crossing, with an energy which falls below the threshold  of the continuum, 
$E_{\rm TH}$. Using two different numerical approaches (finite differences and a mode expansion) they 
estimated the ratio between the ground state energy and the threshold energy, $E_1/E_{\rm TH} \approx 0.66$. 
They also identified a second bound state, with odd-odd symmetry, corresponding to a ratio $E_2/E_{ŧh} \approx 3.72$. 
The energy of this second state falls below the threshold of the continuum for odd-odd states. 
These results were later confirmed by Avishai et al.\cite{Avishai91}. 
More recently, Amore, Fern\'andez and Rodriguez\cite{Amore11},  have used the Conformal Collocation method (CCM)\cite{Amore08} 
 to obtain a precise value for the ground state of the infinite cross, 
using a non uniform grid. They obtained $E_1/E_{\rm TH} \approx 0.659611$. Trefethen and Betcke\cite{Trefethen06}, on the other hand have
obtained a precise value for the second bound state, $E_2/E_{\rm TH} \approx 3.71648$ (notice that
these authors report the value of $E_2$, from which the ratio can be obtained).

Different geometries are closely related to the symmetric cross. Bulgakov and collaborators have studied the properties of 
a configuration of non-orthogonal (``scissor-shaped") crossed wires, observing the emergence of multiple bound 
states below the continuum as the angle between the arms is reduced~\cite{Bulgakov02}.
A particular case of T-shaped waveguide is  obtained
from the symmetric cross
by desymmetrizing the region for even-odd modes, although
this particular system does not support bound states.
Nazarov~\cite{Nazarov10}, analyzed the bound states of a similar configuration, but with arms of
different widths.

Taking into account the above mentioned relevance of the symmetries for the spectrum,
in this paper we study a different modification of the case discussed by Schult et al.\cite{Schult89}, 
considering an asymmetric cross, with orthogonal arms of different width. 
Our goal is to investigate the behavior of the bound states of this system, as one of the arms is enlarged,
and the possibility of the disappearance of these states or the appearance of new bound states. 
The related L and T shaped configurations are also analyzed. 

The paper is organized as follows: in Section \ref{sec_1} we  describe the system of the 
asymmetric cross and we make qualitative predictions on the behavior of the bound states as one of the
arms is enlarged; in Section \ref{results} we present the numerical results, which have been obtained using a 
collocation method; finally, in Section \ref{conclusions} we draw our conclusions.

\section{The asymmetric cross}
\label{sec_1}

As we have mentioned in the Introduction, Schult, Ravenhall and Wyld\cite{Schult89} have proved
that an electron which moves freely in an infinite symmetric cross is localized in the central region 
of the cross when it finds itself in the ground state. They also discovered a second localized state, 
which is the lowest energy state with odd-odd symmetry. 

\begin{figure}[ht]
\begin{center}
\includegraphics[width=5cm]{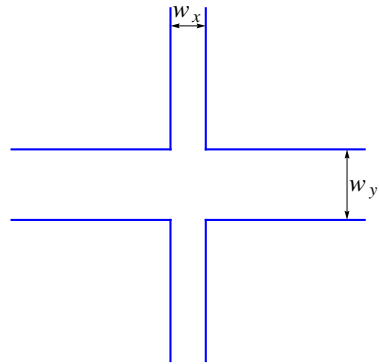}
\caption{(color online) Asymmetric cross  }
\bigskip
\label{fig_1}
\end{center}
\end{figure}

The goal of our paper is to investigate the behavior of these modes (as well as the possible appearance of 
further localized states), as the width of one of the two arms is changed, thus obtaining the asymmetric cross, 
shown in Fig. \ref{fig_1}. 
In particular, we ask ourselves whether the localized modes of the symmetric cross
can survive to arbitrary perturbations of this kind: what happens when one of the arms is much smaller than the
other? Would we still have bound states? What does this imply for the L and T shaped configurations?

To answer our questions we need to solve the scalar Helmholtz equation
\beq
- \frac{1}{2} \ \Delta \Psi_n(x,y) = E_n \Psi_n(x,y) \ ,
\label{eq_1}
\eeq
where $(x,y) \in \Omega = \left\{ |x| < w_x/2 , |y| < w_y/2 \right\}$ and we are using $\hbar/m=1$. 
The wave functions obey 
Dirichlet boundary conditions on $\partial \Omega$.
We have called $w_x$ and $w_y$ the widths of the two arms, defining $\beta \equiv w_y/w_x$ (for $\beta=1$ 
one recovers the symmetric cross). 

Before trying to answer our question in a quantitative way, we may attack the problem qualitatively.
Let us consider the case $\beta \gg 1$, meaning that the width of the horizontal arm is much larger than
the width of the vertical arm, and assume that there is at least one localized solution to eq.(\ref{eq_1}).
This situation is represented in fig.\ref{fig_2}: 
\begin{figure}[ht]
\bigskip
\begin{center}
\includegraphics[width=5cm]{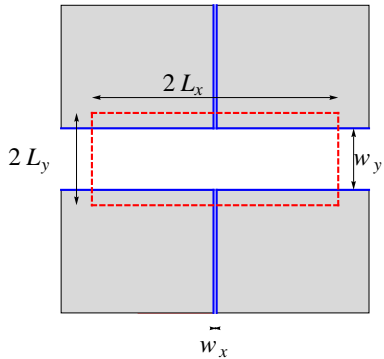}
\caption{(color online) Highly asymmetric cross: $w_x \ll w_y$. }
\bigskip
\label{fig_2}
\end{center}
\end{figure}
the grey regions in this plot represent the forbidden 
region, where the eigenfunctions of eq.~(\ref{eq_1}) must  vanish identically; 
we assume that the dimensions of the red rectangle are large enough  so that the localized wave functions 
are exponentially small outside the rectangle. 

Under these assumptions, one may attempt an approximate solution of the original problem inside the rectangle, and  thus
obtain accurate approximations to the bound states of eq.~(\ref{eq_1}), if they exist. 

The fundamental idea is to represent eq.~(\ref{eq_1}) as a Schr\"odinger equation
\beq
\left[- \frac{1}{2} \Delta + V(x,y) \right] \Psi_n(x,y) = E_n \Psi_n(x,y) \ ,
\label{eq_2}
\eeq
with 
\beq
V(x,y) &=& \left\{ \begin{array}{ccc}
0 & , & (x,y)\in \Omega \\
\infty & , & (x,y) \not\in \Omega
\end{array}\right. \ .
\eeq

To allow numerical calculation the exact potential $V(x,y)$ is replaced by an approximate potential $\tilde{V}(x,y)$:
\beq
\tilde{V}(x,y) &=& \left\{ \begin{array}{ccc}
0 & , & (x,y)\in \Omega \\
V_0 & , & (x,y) \not\in \Omega
\end{array}\right. \ ,
\eeq
which takes a large but finite value $V_0$ in the forbidden region. If $V_0$ is large enough (much larger than the energies 
of the lowest states), the wave functions of these states are exponentially suppressed in the classically forbidden region 
(the grey region in the figure), and the contribution from this region becomes negligible.

The advantage of reformulating the problem in this way should be clear, since we may now work on the entire rectangle, 
$(x,y) \in \mathcal{D}  = \left\{|x| \leq L_x, |y| \leq L_y\right\}$, and use the orthonormal basis 
\beq 
\Phi_{n_x,n_y}(x,y) &=& \chi_{n_x}(x) \ \zeta_{n_y}(y) \nonumber \ ,
\label{basis}
\eeq
with 
\beq
\chi_{n_x}(x) &=& \sqrt{\frac{1}{L_x}} \ \sin \frac{n_x \pi (x+L_x)}{2L_x} \ , \nonumber \\
\zeta_{n_y}(x) &=& \sqrt{\frac{1}{L_y}} \ \sin \frac{n_y \pi (y+L_y)}{2L_y} \ . \nonumber 
\eeq
and $n_x, n_y = 1,2, \dots$.

Kaufman, Kosztin and Schulten~\cite{Kaufman99} have used this approach to calculate approximations
to the eigenvalues and eigenfunctions of the negative Laplacian on finite domains, applying the Rayleigh-Ritz method 
to the Schr\"odinger equation with $V(x,y) \rightarrow \tilde{V}(x,y)$ on a rectangular domain fully enclosing the
region. 

The accuracy of this method clearly depends both on the finite value of the potential step $V_0$ used in the numerical 
calculation and the ratio between the areas of $\mathcal{D}$ and the portion of area of $\Omega$ falling inside $\mathcal{D}$
(which we call $\tilde{\Omega}$): if the area of $\mathcal{D}$ is much
larger than the area of $\tilde{\Omega}$, a large number of basis functions will be needed in the calculation to achieve a suppression 
of the wavefunctions in the classically forbidden regions.

Since we are interested in the limit $L_x \gg L_y$, we may use a simpler and more direct approach: provided that the rectangle 
$\mathcal{D}$ is very thin, the lowest modes of eq.(\ref{eq_2}) on $\mathcal{D}$ will be dominated by the longitudinal modes of the basis, 
with negligible contributions from the transverse excited modes. 

This observation justifies the possibility of using a portion of the Hilbert space of the problem, corresponding to the states
$\Phi_{n_x,1}(x,y)$.

In this limit the ``high--energy" modes essentially decouple from the problem and we may work with an effective Hamiltonian, obtained by
averaging the original Hamiltonian with respect to the mode $\zeta_1(y)$:
\beq
\hat{H}_{eff}(x) &\equiv& \int_{-L_y}^{L_y} dy \ \zeta_1(y) \hat{H} \ \zeta_1(y) \nonumber \\
&=& - \frac{1}{2} \frac{\partial^2}{\partial x^2} + \frac{\pi^2}{8 L_y^2} + V_{eff}(x) \nonumber \ ,
\eeq
where the effective potential is 
\beq
V_{eff}(x)  &\equiv&  \int_{-L_y}^{L_y} dy \ \zeta_1^2(y) \tilde{V}(x,y) \nonumber \\
&=& \left\{ \begin{array}{ccc}
0 & , & |x| \leq w_x/2 \\
2 V_0 \int_{w_y/2}^{L_y} \zeta_1(y)^2 dy & , & |x| > w_x/2 \\
\end{array}\right.
\eeq

\begin{figure}[ht]
\bigskip
\begin{center}
\includegraphics[width=4cm]{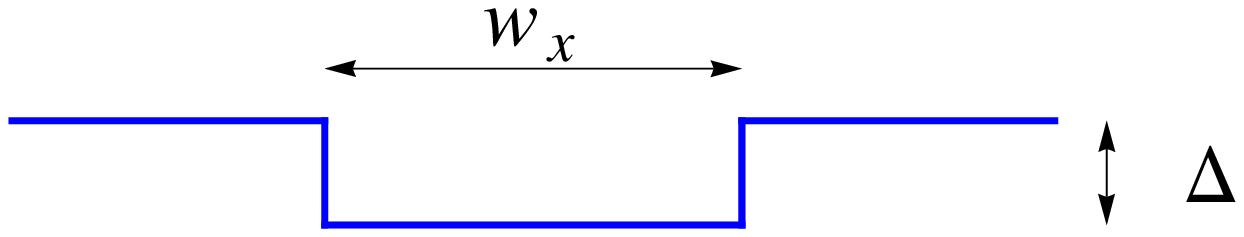}
\includegraphics[width=4cm]{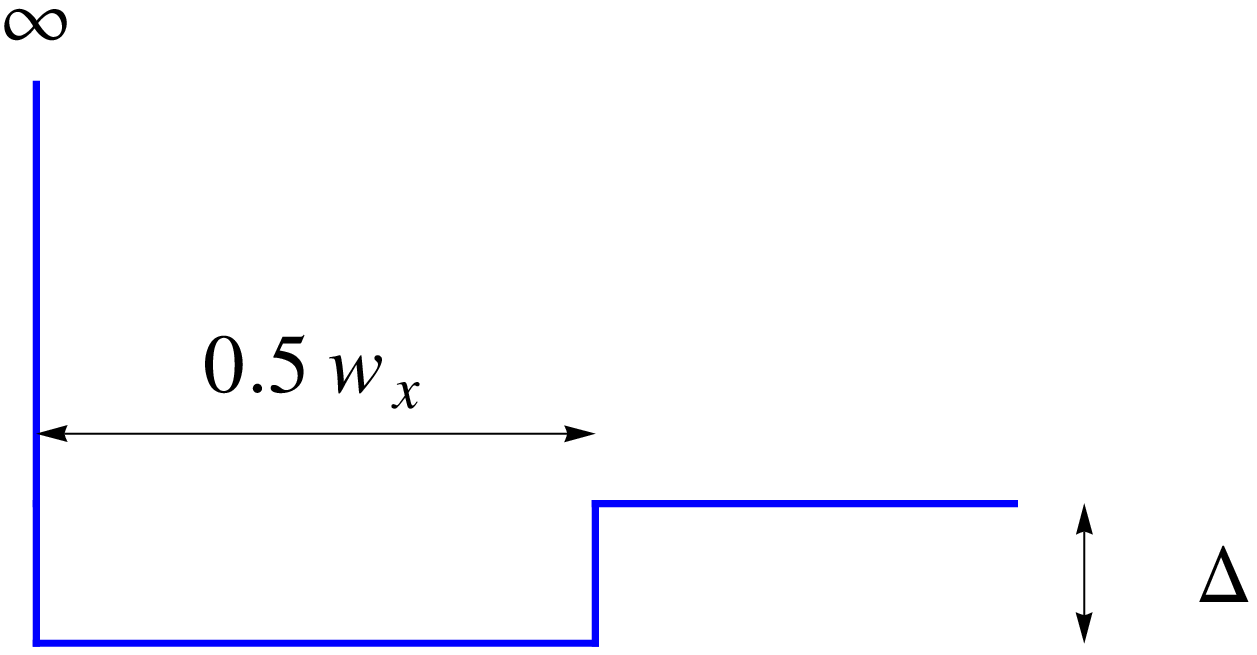}
\caption{(color online) Effective potentials corresponding to figs. \ref{fig_2} and \ref{fig_4} (left) and to figs.\ref{fig_3} and \ref{fig_5} (right)}
\bigskip
\label{fig_6}
\end{center}
\end{figure}

With this qualitative argument we find that the effective potential behaves as a square well of finite depth 
(which we will denote by $\Delta$), 
whose width is precisely the 
width of the smaller arm, $w_x$ (see the left plot of fig.\ref{fig_6}). 
It is well known that the ground state 
of this potential {\sl is always bound}, independently of the width and depth of the well. On these grounds we expect that {\sl the fundamental 
mode of the asymmetric cross of fig.\ref{fig_2} will always be bound, regardless of the width of the two arms, as long as they are finite!}

\begin{figure}[ht]
\bigskip
\begin{center}
\includegraphics[width=4cm]{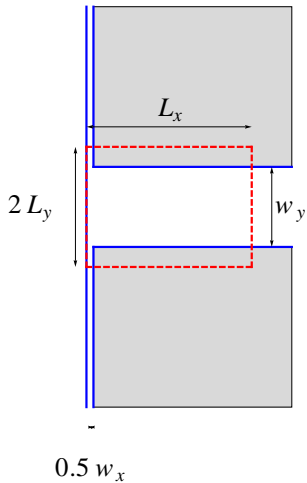}
\caption{(color online) T-shaped configuration; desymmetrized region for odd-even modes of the asymmetric cross. }
\bigskip
\label{fig_3}
\end{center}
\end{figure}

\begin{figure}[ht]
\bigskip\bigskip
\begin{center}
\includegraphics[width=7cm]{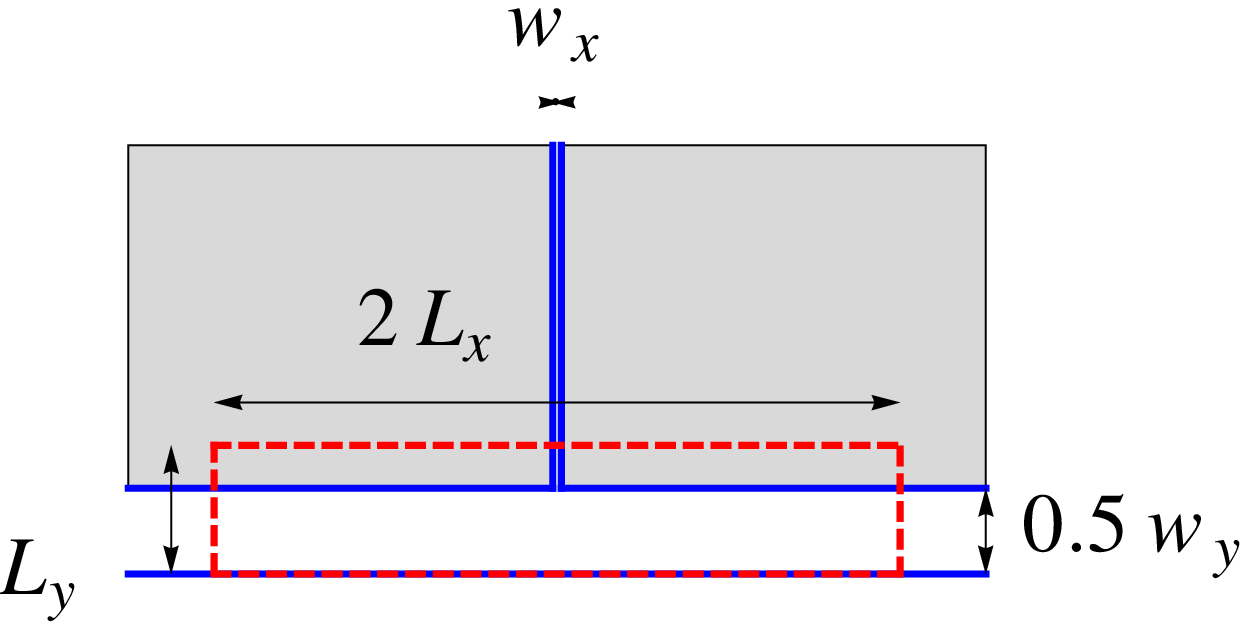}
\caption{(color online)  T-shaped configuration; desymmetrized region for even-odd modes of the asymmetric cross. }
\bigskip
\label{fig_4}
\end{center}
\end{figure}

\begin{figure}[ht]
\bigskip\bigskip
\begin{center}
\includegraphics[width=7cm]{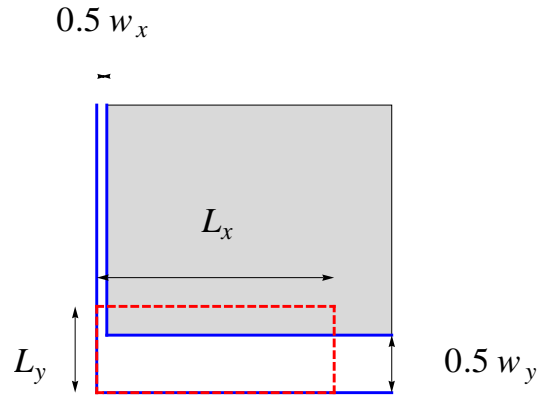}
\caption{(color online) L-shaped configuration; desymmetrized region for odd-odd modes of the asymmetric cross.}
\bigskip
\label{fig_5}
\end{center}
\end{figure}

We can also study the remaining lowest modes in each symmetry class using the same approach. In Figs.\ref{fig_3}, \ref{fig_4}
and \ref{fig_5} we display the desymmetrized regions corresponding to the symmetry classes of modes with 
odd-even, even-odd and odd-odd respectively. Repeating the same steps followed for the full domain of fig.\ref{fig_2}, we may easily 
convince ourselves that the effective potential of Figs.\ref{fig_3} and \ref{fig_5} corresponds to a finite well potential 
with a infinite wall on one side (see the left plot of fig.\ref{fig_6}), whereas the case of Fig.\ref{fig_4} is analogous to the one of Fig.\ref{fig_2}.

~
~

The simple quantum mechanical problem for the potential well of the right plot of fig.\ref{fig_6} {\sl does not}
admit localized solutions when the potential well is either very narrow or very shallow.

These simple arguments allow us to make the following predictions: 
\begin{itemize}
\item  the fundamental mode of the symmetric cross stays always bounded, as the width of one of the two arms is increased arbitrarily, but remains finite; 
\item the fundamental mode of a domain with the shape of an infinite L (which is the lowest energy odd-odd mode of the symmetric cross) 
is bounded, but becomes unbounded as the width of any of the two arms is increased, while keeping fixed the width of the other arm;
\item the fundamental mode of an infinite T shaped domain (even-odd modes of the asymmetric cross), shown in Fig.\ref{fig_4}, is not bound 
for the symmetric cross, but it becomes bound as the width $w_x \rightarrow 0$, but remains finite; 
\item the fundamental mode of the ``rotated" T of fig.~\ref{fig_3} (odd-even modes of the asymmetric cross), is not bound either
for the symmetric cross or as  $w_x \rightarrow 0$.
\end{itemize}

Notice that the predictions regarding the T-shaped infinite domain in Fig.\ref{fig_4} are in agreement with the findings of Nazarov \cite{Nazarov10}.

In the next Section we describe the numerical methods that we have used to test these predictions and we 
present the numerical results obtained with these methods.

\section{Numerical results}
\label{results}

The correctness of the analysis in the previous section strongly depends on the assumption of the decoupling of the ``high-energy" modes
which leads to an effective one-dimensional problem.
However, we are interested in any perturbation of the symmetry,
not only the large ones given by $\beta \gg 1$.
Thus, here we verify, by numerically solving eq.~(\ref{eq_1}), 
that the results obtained in the previous section are correct, 
and, even more, that our predictions can be extrapolated to the whole range of possible values of $\beta$.

We now illustrate the method that we have used in our calculations.
Upon a simple rescaling of the $y$ axis, defining $y' = y/\beta$ ($x' = x$), we write eq.(\ref{eq_1}) in the form of a 
modified Helmholtz equation 
\beq
- \left( \frac{\partial^2}{\partial {x'}^2} + \frac{1}{\beta^2} \frac{\partial^2}{\partial {y'}^2}\right) \psi_n(x',y') = E_n \psi_n(x',y') \ ,
\label{eq_3}
\eeq
which allows us to work on an infinite symmetric cross of width $w_x$,
with $\beta$ playing a role of a parameter. 

As we are interested in the bound states of eq.~(\ref{eq_3}), we may ``cut" the cross at some large but finite distances $L_x$ and $L_y$,
by imposing Dirichlet boundary conditions at $x= \pm L_x$ and $y=\pm L_y$. If $L_x$ and $L_y$ are large enough the wave functions 
of the bound states are exponentially small at larger distances and the error introduced by this approximation is negligible 
compared to the error due to the discretization. In general, for $\beta \gg 1$, one may choose $L_x \gg L_y$, as the probability density is
mostly concentrated on the wider arm. The maximum size of $L_x$ and $L_y$  is essentially dictated by the total number of collocation
points corresponding to this choice. 

On the other hand, excessively large values of $L_x$ and $L_y$ should be avoided, since they
increase the number of collocation points and therefore the computational power needed in the calculation.

To discretize eq.~(\ref{eq_3}) we use a collocation approach based on ``tent functions" (TF): on the interval $|x| \leq L$ we define the
uniform grid, whose $N-1$ points are $x_k \equiv {2Lk}/{N}$, where $N$ is an even integer and $k = -N/2+1 , -N/2+2, \dots, N/2-1$.
The tent function peaked at the point $x_k$ is defined as:
\beq
\phi_k(x) &\equiv&  \left\{ \begin{array}{ccc}
\frac{x-x_{k-1}}{x_k-x_{k-1}} &  , & x_{k-1} \leq x \leq x_k \\
\frac{x_{k+1}-x}{x_{k+1}-x_{k}} &  , & x_{k} \leq x \leq x_{k+1} \\
0 & , & x< x_{k-1} , x>x_{k+1} \\
\end{array}\right.
\eeq

A function $f(x)$ obeying Dirichlet boundary conditions at $x = \pm L$ ($f(\pm L)=0$) can be interpolated using the TF as
\beq
f^{(TF)}(x) = \sum_{k=-N/2+1}^{N/2-1} f(x_k) \phi_k(x) \approx f(x) \nonumber \ .
\eeq

For a higher dimensional problem, the same procedure may be followed, using multidimensional TFs which are the direct product 
of the TFs along each orthogonal direction:
\beq
\Phi_{k_1,k_2, \dots, k_d}(x_1,x_2,\dots , x_d) &=& \phi_{k_1}(x_1) \ \phi_{k_2}(x_2) \ \dots \ \phi_{k_d}(x_d) \nonumber 
\eeq
In principle the interpolation of a d-dimensional function on a d-dimensional region $|x_i| \leq L_i$, requires  
$M = \prod_{i=1}^d (N_i-1)$ points.

The eigenvalue equation (\ref{eq_3}) may then be converted to a matrix eigenvalue problem calculating the matrix elements of the Hamiltonian 
operator between all the functions of a set: the size of the matrix depends on the total number of functions used in the calculation,
which rapidly grows with the dimensionality of the problem. 
A drastic reduction of the number of functions however can be achieved by considering only functions which are peaked at points 
internal to the cross. 

A second observation concerns the choice of $N_i$, i.e. the number of collocation points in each orthogonal 
direction: although in principle, $N_i$ ($i=1,\dots d$) can take any integer value, it is convenient to pick values of $N_i$ which 
allow to sample exactly the border of the cross. The eigenvalues obtained 
for different grids sampling the border of the domain provide a monotonous sequence of values, which allows to obtain 
a good approximation via extrapolation\cite{Amore10e}.

Before calculating the eigenvalues of the asymmetric cross, we have tested this collocation method on the symmetric cross, for which a
precise result is available~\cite{Amore11}. The method used in that case is the conformal collocation method (CCM)~\cite{Amore08}.
Working with grids with $N = N_x = N_y = 80, 120, \dots, 880$ we have calculated the lowest eigenvalue of the corresponding collocation matrix, 
obtaining a monotonous sequence of values: the most accurate value, corresponding to the finest grid, provides
a ratio $E_1/E_{\rm TH} = 0.66166$, which is just $0.3 \%$ above the value reported by Amore et al.~\cite{Amore11} 
We have also performed a least square fit 
of the monotonous sequence of values with the functional form $a_1 + a_2/N^\gamma + a_3/N^{2\gamma}+a_4/N^{3\gamma}$, obtaining
$a_1 = 0.65955$, which differs just for the $0.1 \%$ from the  value previously reported\cite{Amore11}. 
This test makes us confident that our 
results are precise.

We can now present the results obtained for the asymmetric cross, discussing separately the states with different symmetries. 
In all cases we expect that the wave function of a bound state will decay exponentially as one moves far away from the center of the cross; 
in particular, the effective Hamiltonian approach that we have described in the previous section predicts $\Psi(x,y)|_{y \rm  \ fixed} \approx e^{-x/\ell_x}$ 
for $|x|\gg w_x/2$, in the limit $\beta \gg 1$. 
We have found that there is at most only one bound state in each symmetry class.
 
\subsection{Even-even state}

We first study the lowest energy state with even-even symmetry of the asymmetric cross. 
In the third column of Table \ref{tab_ee} 
\begin{table}[hbp]
\caption{The ratio $E^{\rm (ee)}/E_{\rm TH}$ for the lowest even-even state of the asymmetric cross. 
Set I corresponds to using $L = 20$ and $N=600$; Set II corresponds to using
$L=40$ and $N=800$; Set III corresponds to using $L=100$ and $N=1600$.}
\label{tab_ee}
\begin{ruledtabular}
\begin{tabular}{|l|l|c|cc|}
$\beta$          & Set & $E^{\rm (ee)}/E_{\rm TH}$ & $\ell_x$ & $\ell_y$ \\
\hline
1.0  & I   & 0.662960 & 1.098   & 1.098 \\
1.1  & I   & 0.723925 & 1.335   & 1.006 \\
1.2  & I   & 0.774665 & 1.613   & 0.938 \\
1.3  & I   & 0.816242 & 1.936   & 0.887 \\
1.4  & I   & 0.849968 & 2.309   & 0.847 \\
1.5  & II  & 0.879058 & 2.770   & 0.818 \\
1.6  & II  & 0.900702 & 3.267   & 0.793 \\
1.7  & II  & 0.918059 & 3.830   & 0.773 \\
1.8  & II  & 0.931999 & 4.463   & 0.757 \\
1.9  & II  & 0.943228 & 5.172   & 0.744 \\
2.0  & II  & 0.952308 & 5.960   & 0.733 \\
2.1  & II  & 0.959682 & 6.832   & 0.723 \\
2.2  & III & 0.965821 & 7.959 & 0.717 \\
2.3  & III & 0.970627 & 9.053 & 0.711  \\
2.4  & III & 0.974578 & 10.252 & 0.705\\
2.5  & III & 0.977844 & 11.564 & 0.700 \\
2.6  & III & 0.980557 & 12.992 & 0.696 \\
2.7  & III & 0.982821 & 14.542 & 0.695 \\
2.8  & III & 0.984719 & 16.219 & 0.690 \\
2.9  & III & 0.986319 & 18.026 & 0.689 \\
3.0  & III & 0.987674  & 19.965 & 0.685 \\
\end{tabular}
\end{ruledtabular}
\bigskip\bigskip
\end{table}
we report the values of the ratio $E^{\rm (ee)}/E_{\rm TH}$, obtained for 
different values $\beta$. We used three different sets, corresponding to  
$L=20$ and $N=600$ (set I), $L=40$ and $N=800$ (set II) and $L=100$ and $N=1600$ (set III).
The fourth and fifth column report the values of the decay lengths $\ell_x$ and $\ell_y$
obtained fitting the numerical eigenfunctions with  the asymptotic behaviors $\Psi(x,0) \approx e^{-x/\ell_x}$ 
and $\Psi(0,y) \approx e^{-y/\ell_y}$, respectively for $|x| \gg w_x/2 $ and $|y| \gg w_y/2$.

\begin{figure}[ht]
\bigskip
\begin{center}
\includegraphics[scale=0.27]{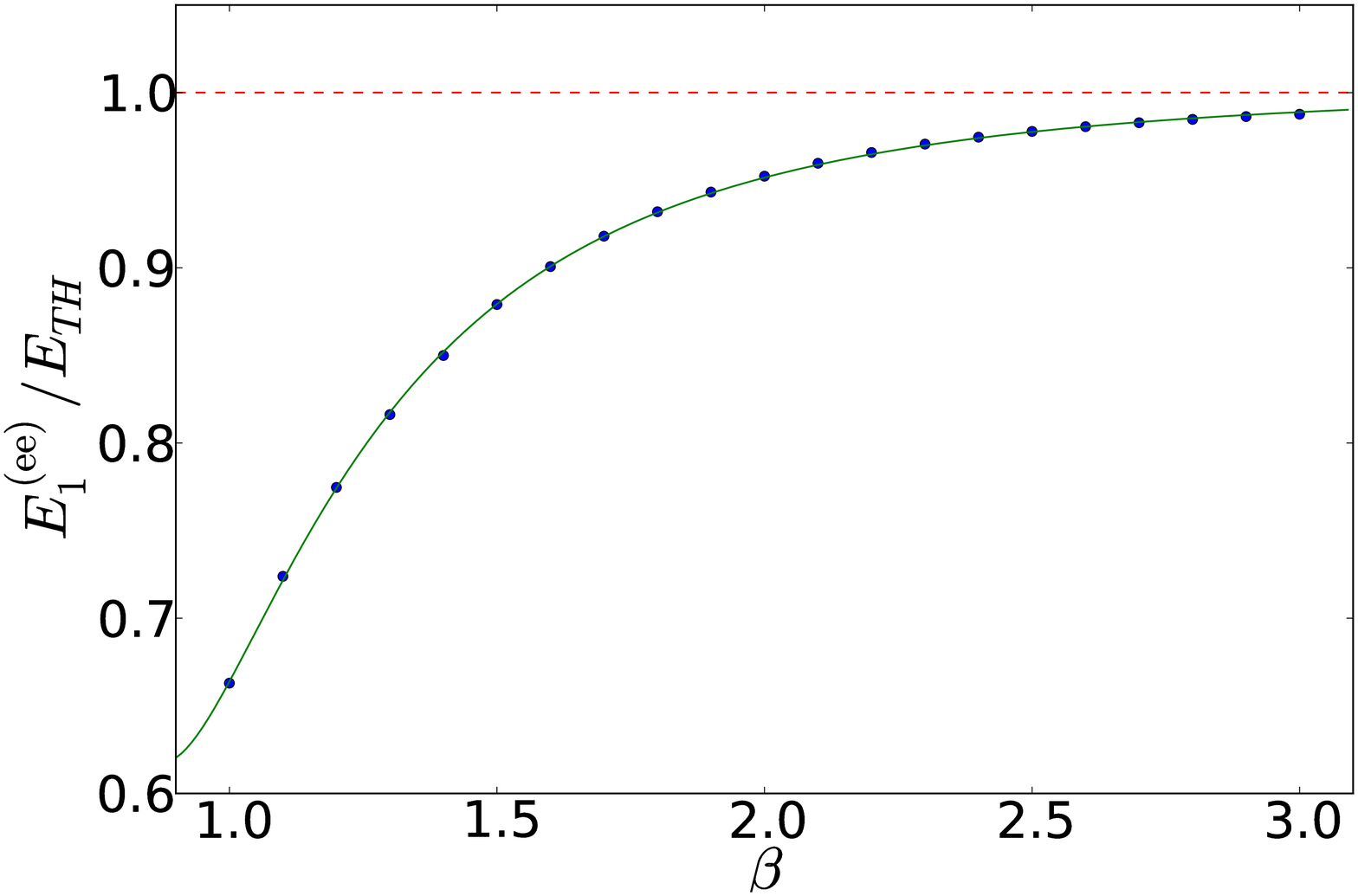}
\caption{(color online) Ratio $E^{(\rm ee)}/E_{\rm TH}$ 
as a function of $\beta$
for the asymmetric cross;  the solid line is the least squares fit 
$\left. E_1/E_{\rm TH} \right|^{\rm FIT} =  1.003 -{0.505}{\beta ^{-3.23725}}+{0.165}{\beta^{-6.475}}$.}
\bigskip
\label{fig_ee1}
\end{center}
\end{figure}

\begin{figure}[ht]
\bigskip
\begin{center}
\includegraphics[scale=0.27]{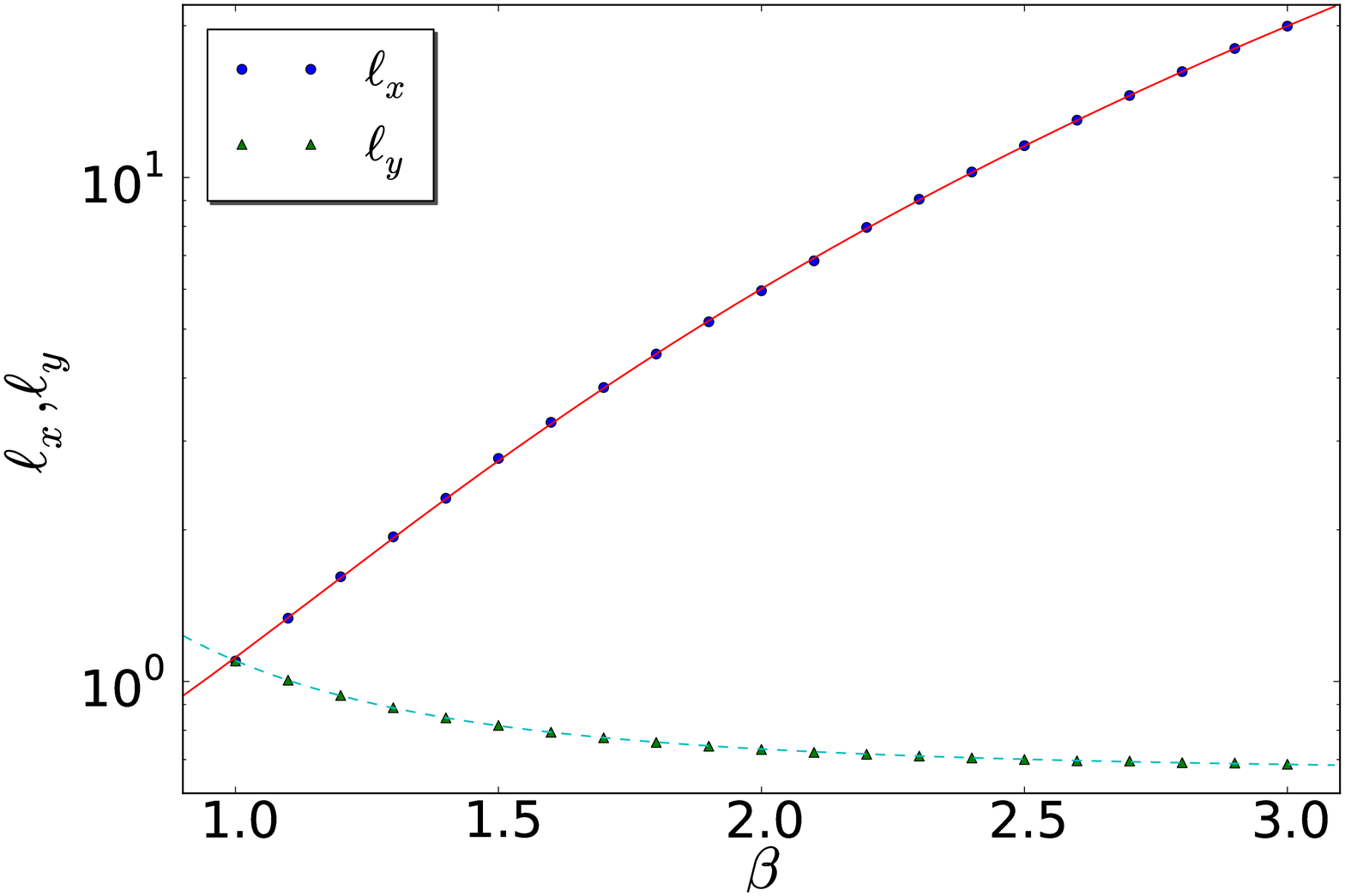}
\caption{(color online) $\ell_x$ and $\ell_y$ 
as a function of $\beta$
for the lowest even-even state of the asymmetric cross; 
the solid and dashed lines are the least squares fits
$\left. \ell_x \right|^{\rm FIT} =  0.472 + 0.644 \ \beta ^{3.104}$ and
$\left. \ell_y \right|^{\rm FIT} =  0.656 + {0.443}{\beta ^{-2.503}}$.}
\bigskip
\label{fig_ee2}
\end{center}
\end{figure}

These values are plotted in Figs.\ref{fig_ee1} and \ref{fig_ee2}. 

The dependence on $\beta$ is described very 
accurately by simple least square fits. In particular, 
within the accuracy of our calculations,
the fit of $E^{\rm (ee)}/E_{\rm TH}$ 
is consistent with the survival of the bound state for $\beta \rightarrow \infty$.

\begin{figure}[ht]
\bigskip
\begin{center}
\includegraphics[scale=0.4]{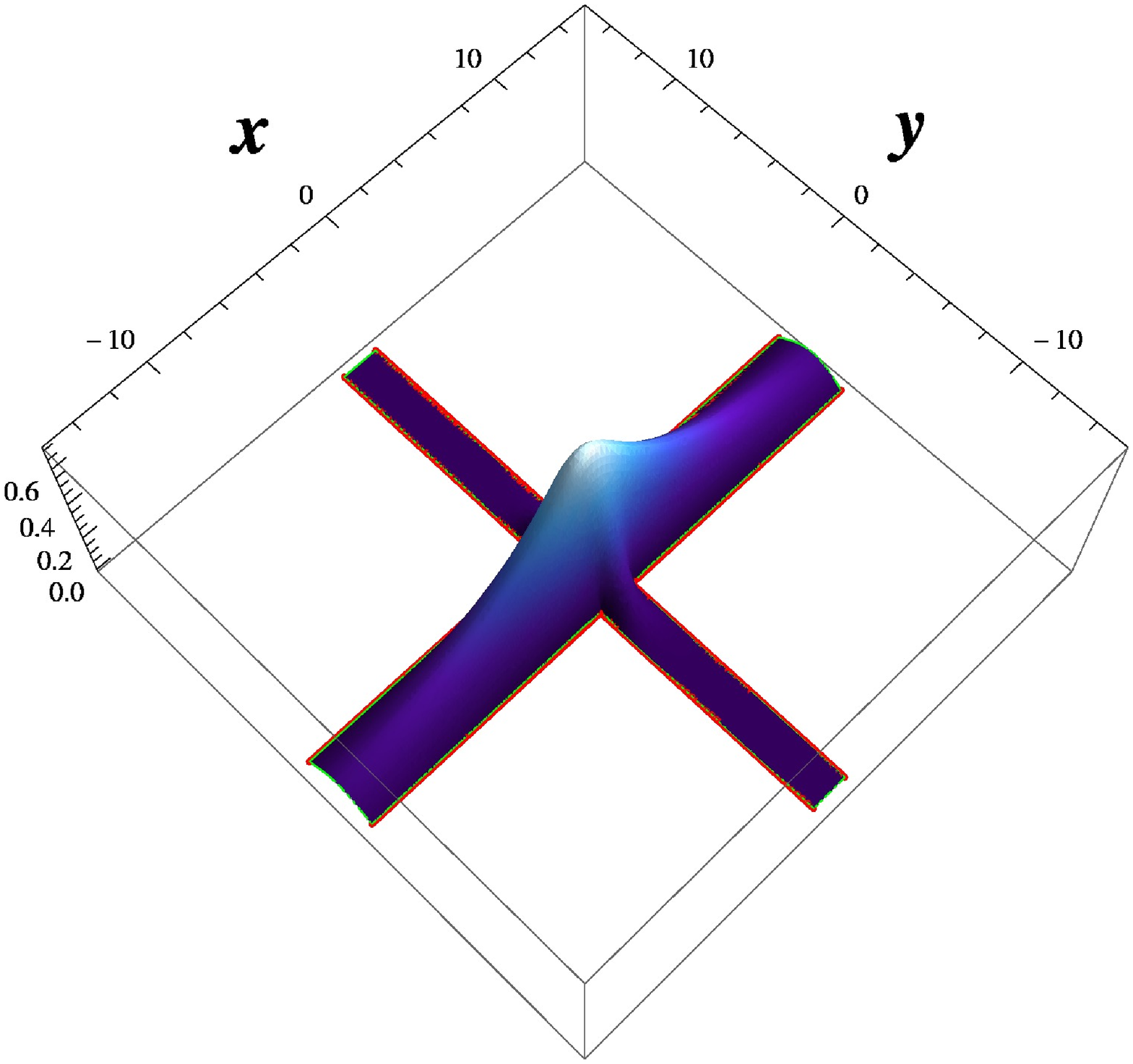}
\caption{(color online) Wave function of the lowest even-even state for $\beta = 2$.}
\bigskip
\label{fig_eewf3D}
\end{center}
\end{figure}

\begin{figure}[ht]
\bigskip
\begin{center}
\includegraphics[scale=0.5]{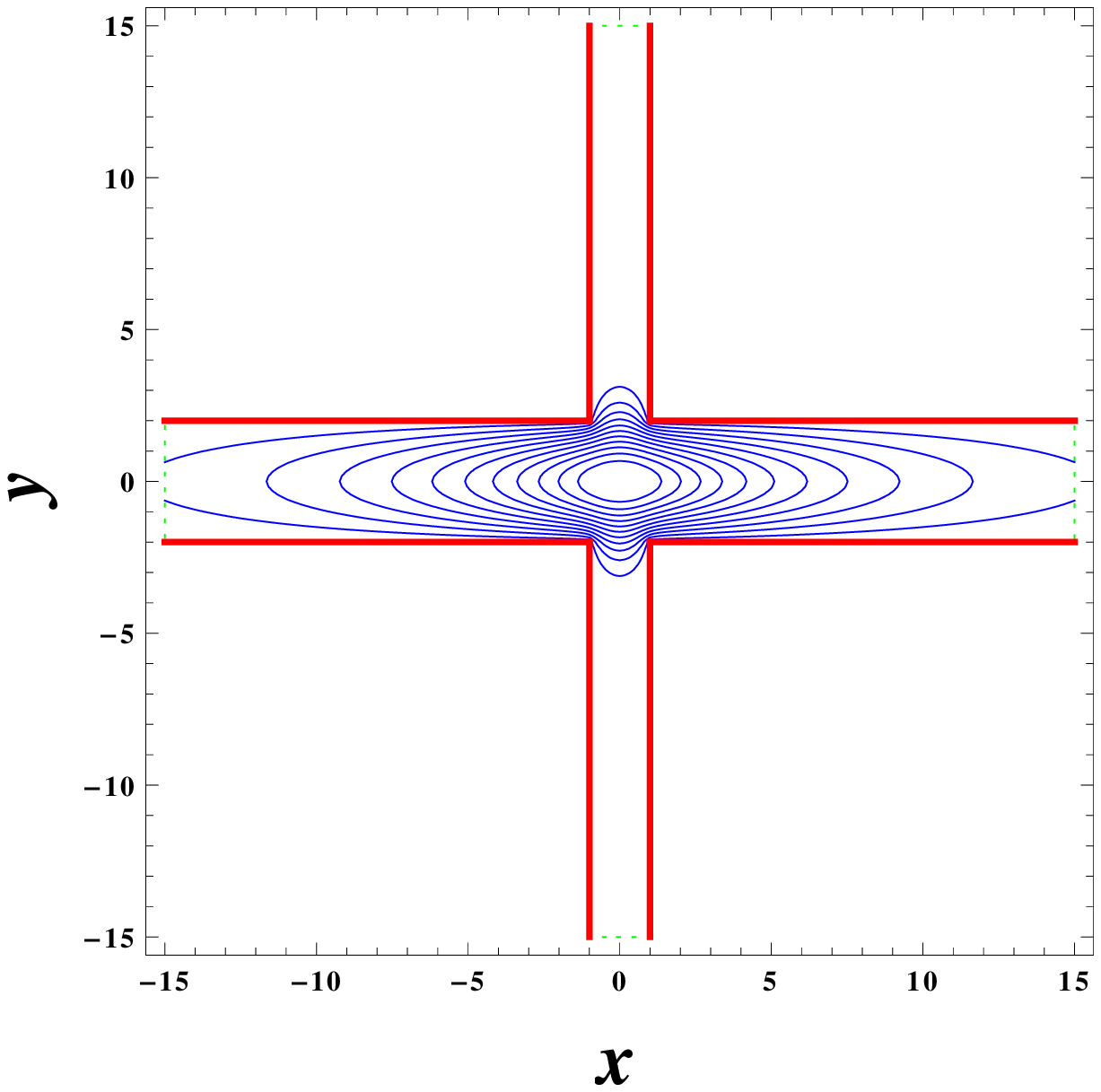}
\caption{(color online) Wave function of the lowest even-even state for $\beta = 2$.}
\bigskip
\label{fig_eewf}
\end{center}
\end{figure}

Notice that $\ell_x$ grows roughly cubically in $\beta$ for $\beta \gg 1$, limiting the range of values of
$\beta$ where the numerical calculation can be performed.
As an example, in figs.\ref{fig_eewf3D} and \ref{fig_eewf} we show different representations of our numerical results
for the wave function of the lowest even-even state for $\beta = 2$ (in these figures, as well in the other figures representing the
wavefunctions, the region of the plot is smaller than the actual region where the numerical calculations were performed).

\subsection{Odd-odd state}

As it was found by Schult, Ravenhall and Wyld\cite{Schult89}  the symmetric cross ($\beta = 1$) has a second bound state, which has odd-odd symmetry and therefore
it is also an eigenstate of an infinite L. According to our qualitative discussion based on an effective Hamiltonian, we expect that
this state may become unbound at some finite $\beta$.

In Table \ref{tab_oo} we report the results for the ratio $E^{\rm (oo)}/E_{\rm TH}$ (second column) and for the 
longitudinal and transverse length scales, $\ell_x$ and $\ell_y$ (third and fourth columns). In this case the length scales are 
obtained fitting the asymptotic behavior of the wave function with $\Psi(x,w_y/3) \approx e^{-x/\ell_x}$ 
and $\Psi(w_y/3,y) \approx e^{-y/\ell_y}$, respectively for $|x| \gg w_x/2 $ and $|y| \gg w_y/2$, since the wave function
vanishes for $x=0$ or $y=0$.

\begin{table}[tbp]
\caption{The ratio $E^{\rm (oo)}/E_{\rm TH}$ for the lowest odd-odd state of the 
asymmetric cross. All the results are obtained using $L=100$ and $N=1600$.}
\label{tab_oo}
\begin{ruledtabular}
\begin{tabular}{|l|c|cc|}
$\beta$  & $E^{\rm (oo)}/E_{\rm TH}$ & $\ell_x$ & $\ell_y$ \\
\hline
1.00  &  3.72042 & 1.332 & 1.332 \\
1.01  &  3.75611 & 1.465 & 1.232 \\
1.02  &  3.78877 & 1.623 & 1.148 \\
1.03  &  3.81839 & 1.816 & 1.076 \\
1.04  &  3.84499 & 2.055 & 1.014 \\
1.05  &  3.86855 & 2.359 & 0.960 \\
1.06  &  3.88909 & 2.760 & 0.912 \\
1.07  &  3.90659 & 3.313 & 0.870 \\
1.08  &  3.92107 & 4.125 & 0.832 \\
1.09  &  3.93252 & 5.429 & 0.797 \\
1.10  &  3.94095 & 7.875 & 0.766 \\
1.11  &  3.94635 & 14.119 & 0.739 \\
1.111 &  3.94673 & 15.322 & 0.735 \\
1.112 &  3.94707 & 16.745 & 0.733 \\
1.113 &  3.94739 & 18.456 & 0.730 \\
1.114 &  3.94767 & 20.549 & 0.728 \\
1.115 &  3.94792 & 23.167 & 0.726 \\
1.116 &  3.94815 & 26.528 & 0.722 \\
\end{tabular}
\end{ruledtabular}
\bigskip\bigskip
\end{table}

\begin{figure}[ht]
\bigskip
\begin{center}
\includegraphics[scale=0.27]{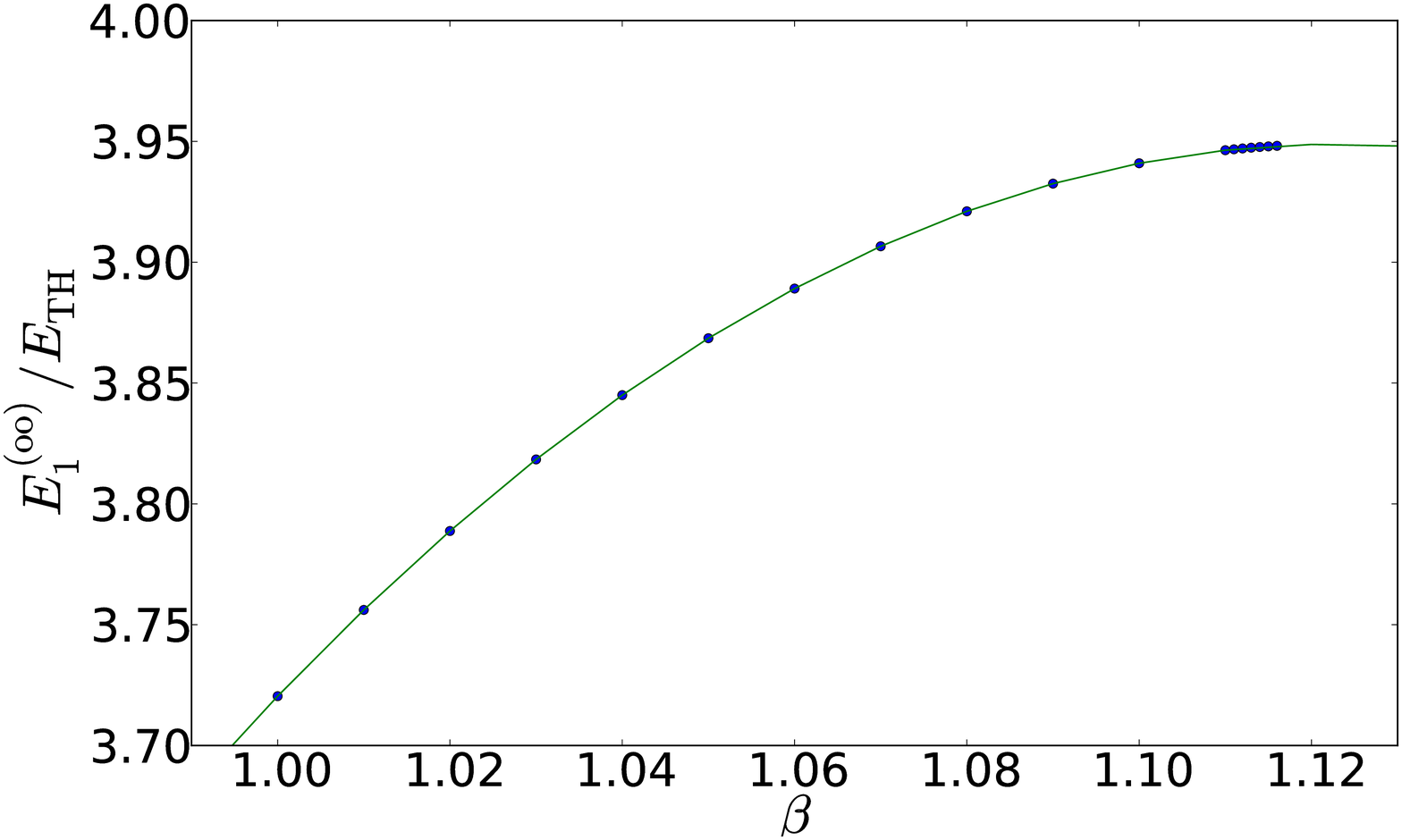}
\caption{(color online) Ratio $E^{(\rm oo)}/E_{\rm TH}$ as a function of $\beta$ 
for the asymmetric cross; 
the solid line is the least squares fit 
$\left. E_1^{({\rm oo})}/E_{\rm TH} \right|^{\rm FIT} = -15.61-15.56 \beta ^{1.974}+34.88 \beta ^{0.9869}$.}
\bigskip
\label{fig_oo1}
\end{center}
\end{figure}

In Fig.\ref{fig_oo1} 
we report the results for $E^{(\rm oo)}/E_{\rm TH}$  of the Table \ref{tab_oo} and compare them with the least squares fit
\[\left. E_1^{({\rm oo})}/E_{\rm TH} \right|^{\rm FIT} = -15.61-15.56 \beta ^{1.974}+34.88 \beta ^{0.9869}\, .\] 
Notice that
the fit has a maximum corresponding to $\beta =  1.12288$.

\begin{figure}[ht]
\bigskip
\begin{center}
\includegraphics[scale=0.27]{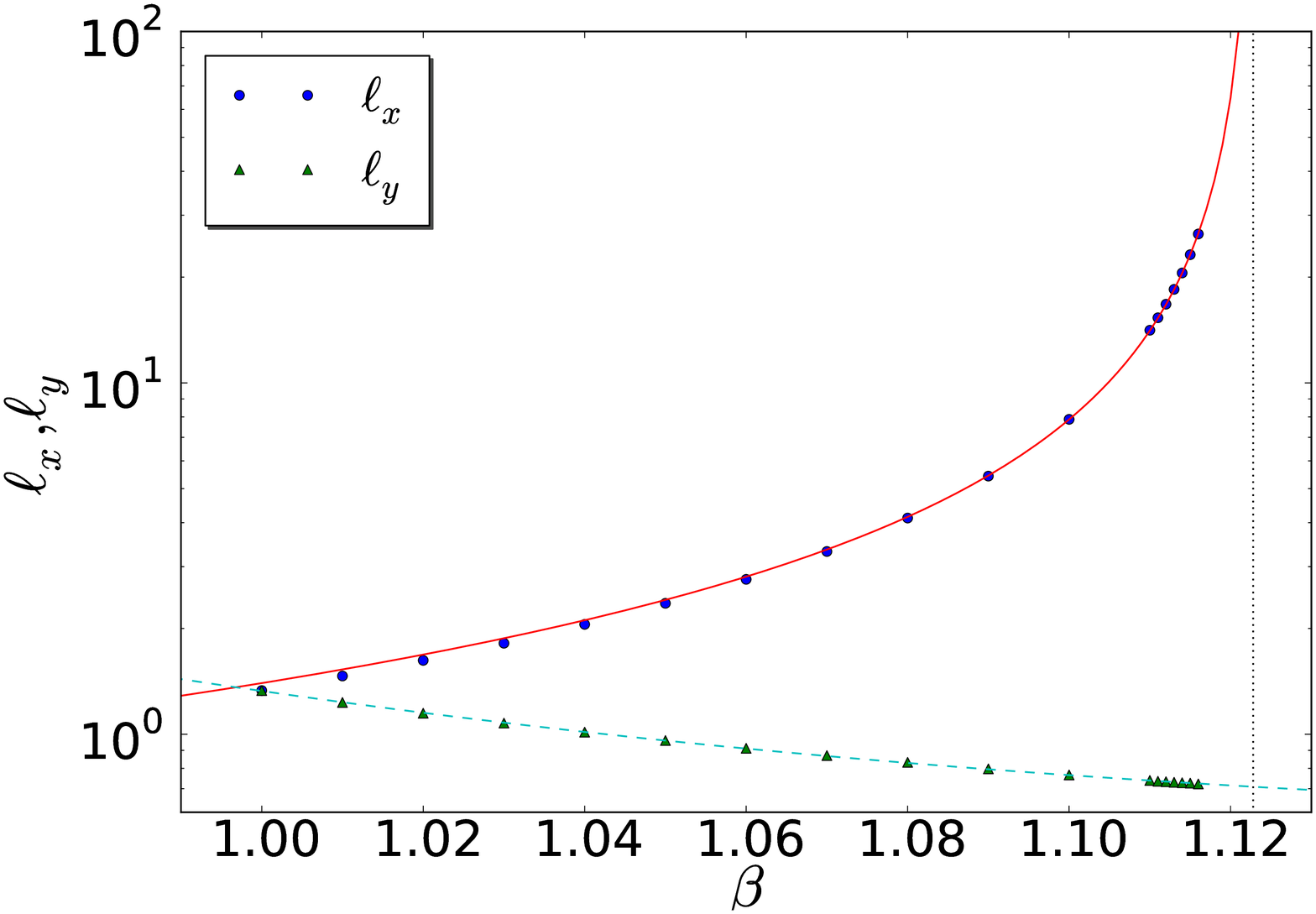}
\caption{(color online) $\ell_x$ and $\ell_y$ as a function of $\beta$
for the lowest odd-odd state of the asymmetric cross; 
the solid and dashed lines are the least squares fits
$\left. \ell_x \right|^{\rm FIT} =  {0.0108614}/({1 -0.992232 \beta ^{0.0673327}})$ and
$\left. \ell_y \right|^{\rm FIT} =  0.514767 +{0.81312}{\beta ^{-12.3632}}$. The vertical line corresponds to
the critical value $\beta^{({\rm oo})}_\star = 1.2279$ where $\left. \ell_x \right|^{\rm FIT}$ is singular.}
\bigskip
\label{fig_oo2}
\end{center}
\end{figure}

\begin{figure}[ht]
\bigskip
\begin{center}
\includegraphics[scale=0.4]{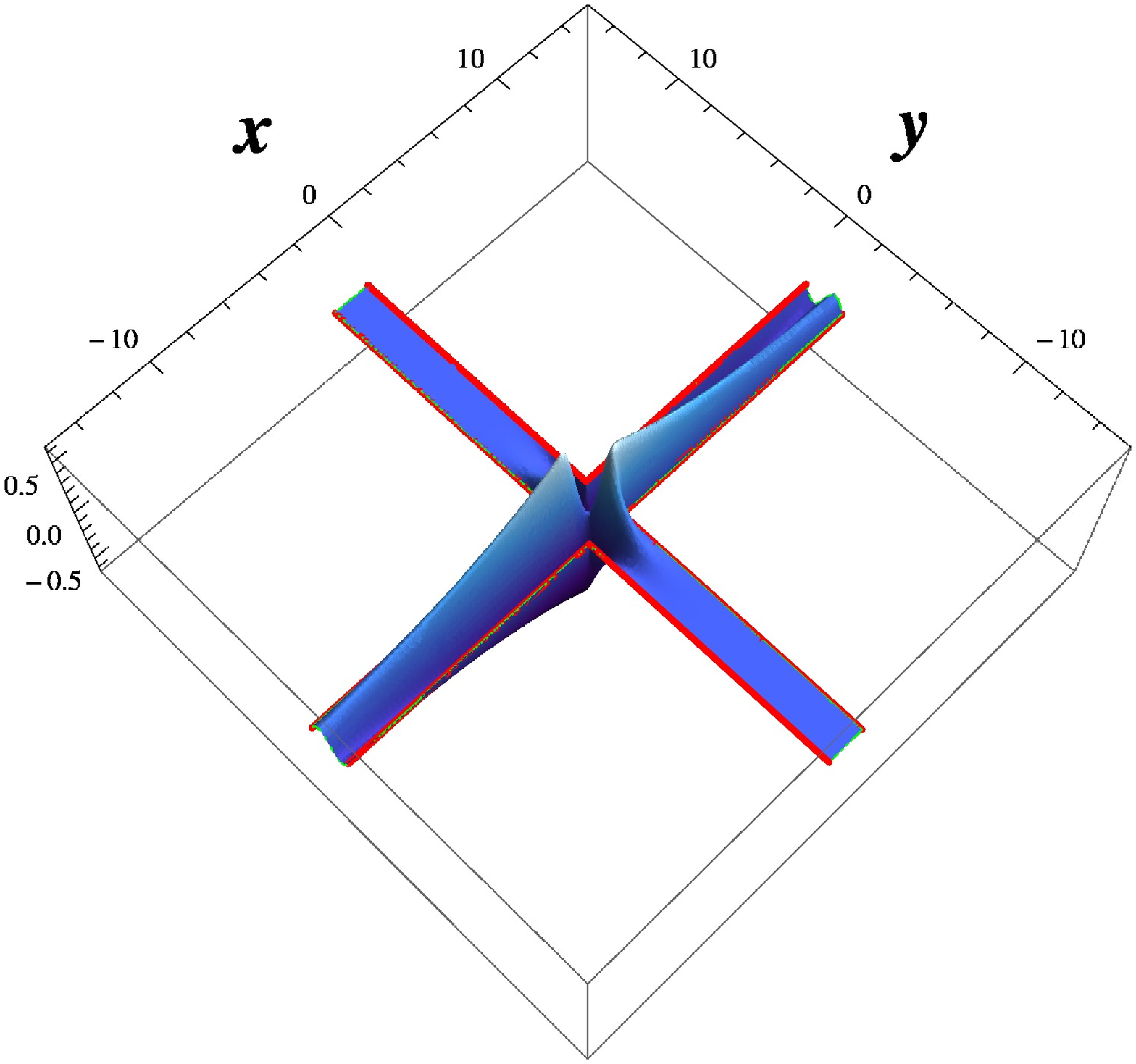}
\caption{(color online) Wave function of the lowest odd-odd state for $\beta = 1.1$.}
\bigskip
\label{fig_oowf3D}
\end{center}
\end{figure}

\begin{figure}[ht]
\bigskip
\begin{center}
\includegraphics[scale=0.5]{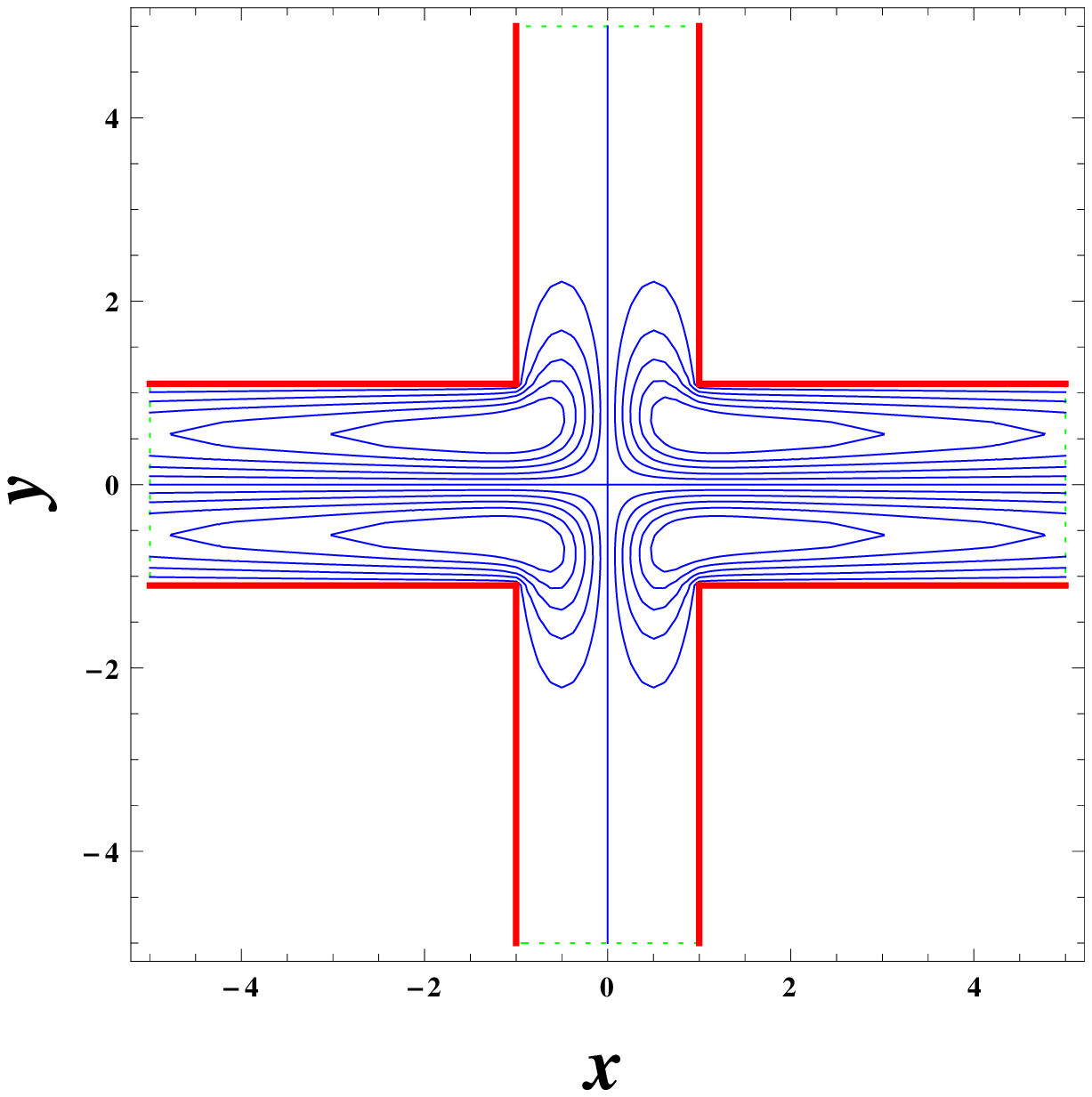}
\caption{(color online) Wave function of the lowest odd-odd state for $\beta = 1.1$.}
\bigskip
\label{fig_oowf}
\end{center}
\end{figure}

In Fig.\ref{fig_oo2} 
we display the length scales $\ell_x$ and $\ell_y$ as functions of $\beta$, and compare them with the fits
\[\left. \ell_x \right|^{\rm FIT} =  \frac{0.0108614}{1 -0.992232 \beta ^{0.0673327}}\] and
\[\left. \ell_y \right|^{\rm FIT} =  0.514767 +\frac{0.81312}{\beta ^{12.3632}}.\] We believe that this plot offers
convincing evidence of the existence of a singularity in $\ell_x$ close to $\beta^{({\rm oo})}_\star = 1.2279$: 
for $\beta \geq \beta^{({\rm oo})}_\star $ the wave function becomes unbound, in agreement with our earlier prediction.
As an example, in figs.\ref{fig_oowf3D} and \ref{fig_oowf} we show different representations of our numerical results
for the wave function of the lowest odd-odd state for $\beta = 1.1$.

\subsection{Even-odd state}

\begin{figure}[ht]
\bigskip
\begin{center}
\includegraphics[scale=0.4]{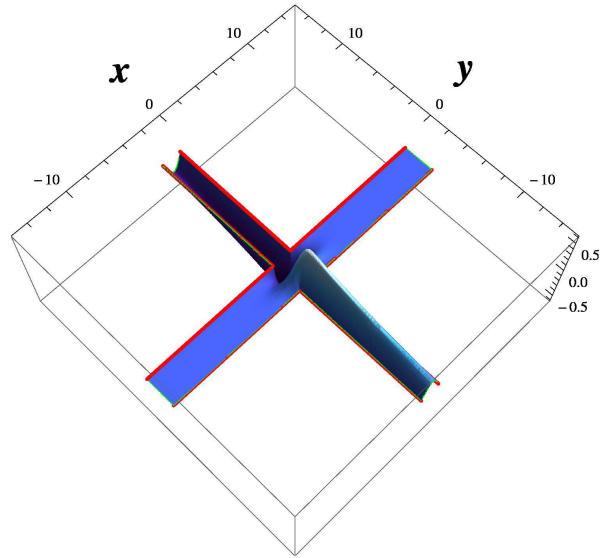}
\caption{(color online) Wave function of the lowest even-odd state for $\beta = 1.55$.}
\bigskip
\label{fig_eowf3D}
\end{center}
\end{figure}

\begin{figure}[ht]
\bigskip
\begin{center}
\includegraphics[scale=0.5]{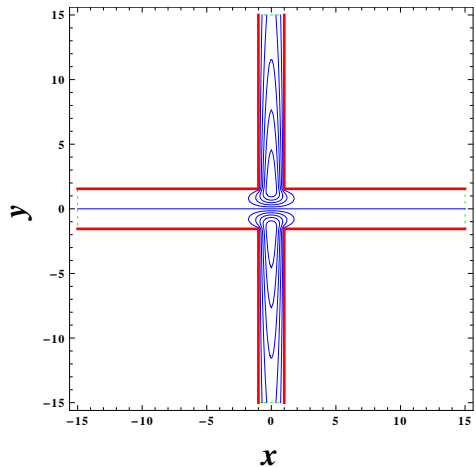}
\caption{(color online) Wave function of the lowest even-odd state for $\beta = 1.55$.}
\bigskip
\label{fig_eowf}
\end{center}
\end{figure}

\begin{figure}[ht]
\bigskip
\begin{center}
\includegraphics[scale=0.4]{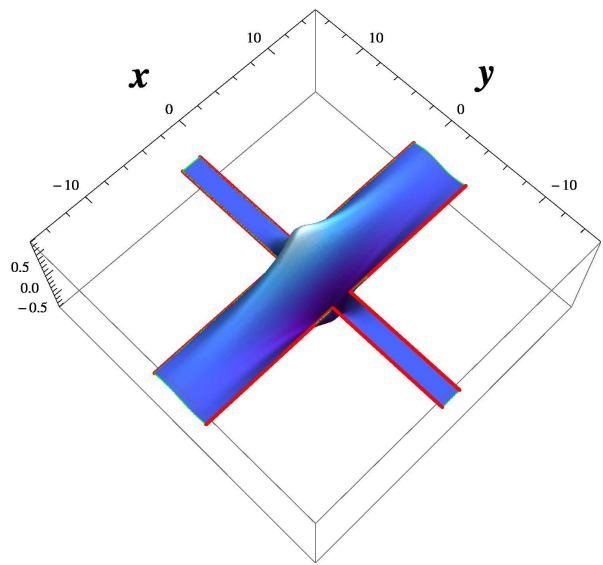}
\caption{(color online) Wave function of the lowest even-odd state for $\beta = 3$.}
\bigskip
\label{fig_eowf3Db}
\end{center}
\end{figure}

\begin{figure}[ht]
\bigskip
\begin{center}
\includegraphics[scale=0.5]{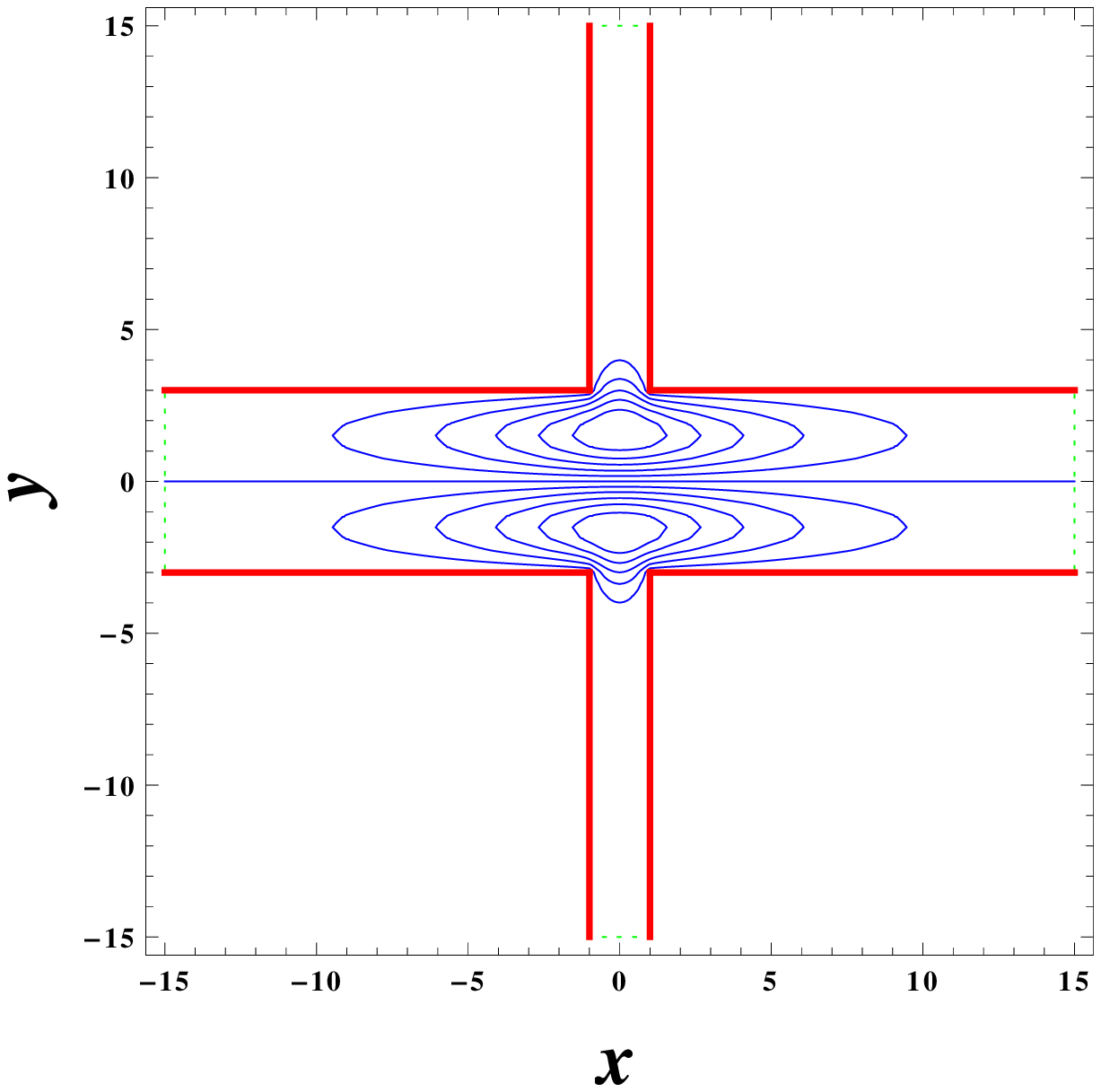}
\caption{(color online) Wave function of the lowest even-odd state for $\beta = 3$.}
\bigskip
\label{fig_eowfb}
\end{center}
\end{figure}

In  Table \ref{tab_eo} 
\begin{table}
\caption{The ratio $E^{\rm (eo)}/E_{\rm TH}$ for the lowest odd-odd state of the 
asymmetric cross. All the results are obtained using $L=100$ and $N=1600$.}
\label{tab_eo}
\begin{ruledtabular}
\begin{tabular}{|l|c|cc|}
$\beta$  & $E^{\rm (eo)}/E_{\rm TH}$ & $\ell_x$ & $\ell_y$ \\
\hline
1.530 &  2.33233 & 0.767 & 29.993 \\
1.531 &  2.33526 & 0.768 & 28.322 \\
1.532 &  2.33817 & 0.769 & 26.830 \\
1.533 &  2.34108 & 0.771 & 25.490 \\
1.534 &  2.34399 & 0.772 & 24.279 \\
1.535 &  2.34689 & 0.773 & 23.179 \\
1.536 &  2.34978 & 0.774 & 22.176 \\
1.537 &  2.35267 & 0.775 & 21.258 \\
1.538 &  2.35555 & 0.777 & 20.415 \\
1.539 &  2.35843 & 0.778 & 19.637 \\
1.54 &  2.36130 & 0.779 & 18.917 \\
1.55 &  2.38974 & 0.791 & 13.879 \\
1.56 &  2.41764 & 0.803 & 10.999 \\
1.57 &  2.44502 & 0.816 & 9.136 \\
1.58 &  2.47190 & 0.828 & 7.832 \\
1.59 &  2.49827 & 0.841 & 6.869 \\
1.6  &  2.52415 & 0.854 & 6.127 \\
1.7  &  2.75775 & 0.992 & 3.087 \\
1.8  &  2.95102 & 1.148 & 2.172 \\
1.9  &  3.11087 & 1.322 & 1.734 \\
2.0  &  3.24315 & 1.516 & 1.478 \\
2.1  &  3.35274 & 1.732  & 1.311\\
2.2  &  3.44367 & 1.971  & 1.195\\
2.3  &  3.51929 & 2.235  & 1.109\\
2.4  &  3.58233 & 2.525  & 1.043\\
2.5  &  3.63503 & 2.842  & 0.992\\
2.6  &  3.67922 & 3.189  & 0.951\\
2.7  &  3.71638 & 3.567  & 0.917\\
2.8  &  3.74774 & 3.977  & 0.890\\
2.9  &  3.77430 & 4.421  & 0.866\\
3.0  &  3.79685 & 4.902  & 0.847\\
4.0  &  3.90424 & 12.077  & 0.747\\
5.0  &  3.93252 & 24.947  & 0.717\\
\end{tabular}
\end{ruledtabular}
\bigskip\bigskip
\end{table}
we report the  results for $E^{\rm (eo)}/E_{\rm TH}$, $\ell_x$ and $\ell_y$ of this state,
calculated using $L=100$ and $N=1600$. The behavior of the lowest even-odd state is rich and it is characterized by
three different behaviors: a first region, below a critical value of $\beta$, where the wave function is unbound; a 
second region, where the wave function is bound, but mainly localized in the vertical arm
(see for instances figs.\ref{fig_eowf3D} and \ref{fig_eowf}), 
and a third region, 
where the wave function is still bound, but mainly localized in the horizontal arm
(see for instances figs.\ref{fig_eowf3Db} and \ref{fig_eowfb}). 

\begin{figure}[ht]
\bigskip
\begin{center}
\includegraphics[scale=0.27]{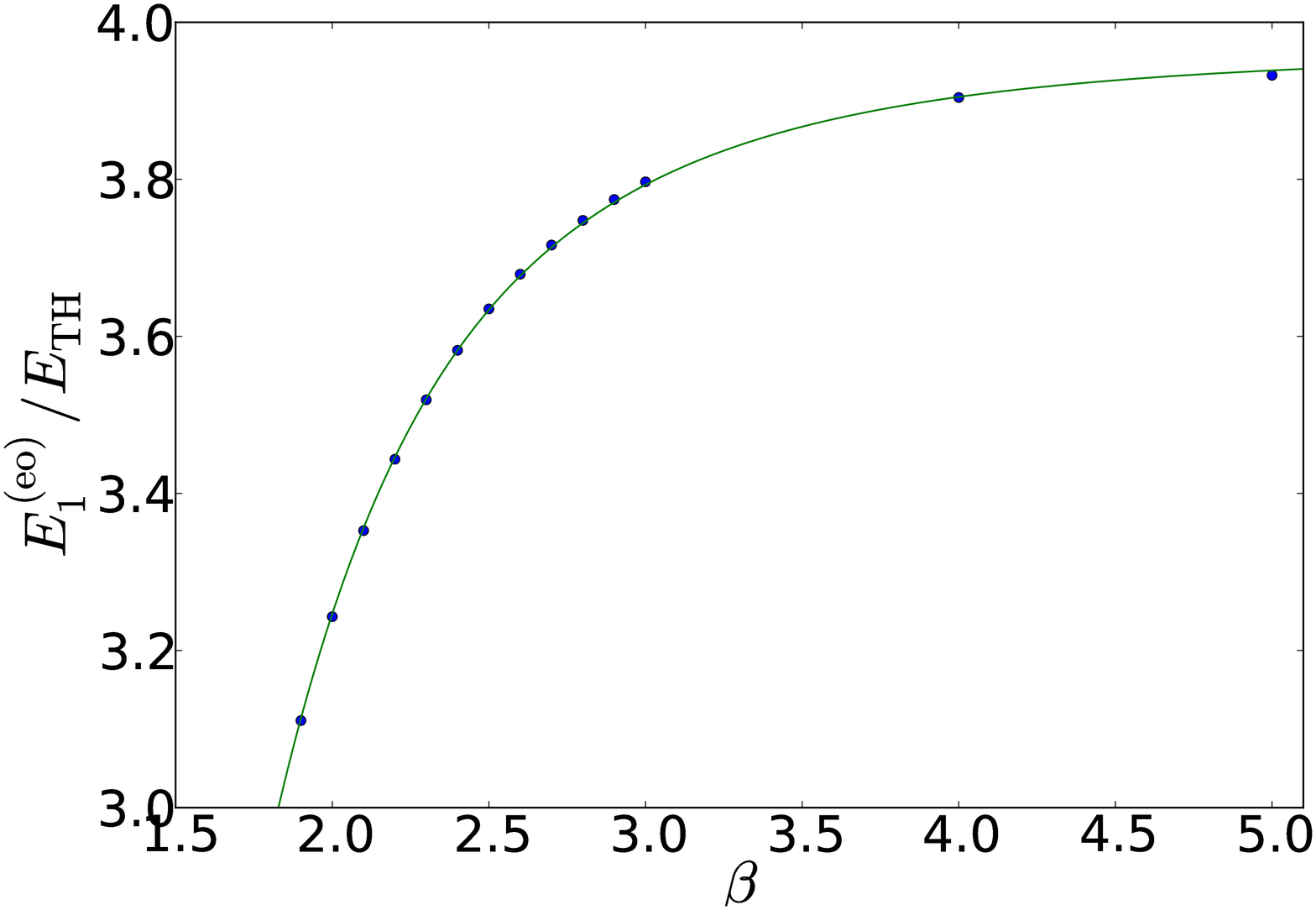}
\caption{(color online) Ratio $E^{(\rm eo)}/E_{\rm TH}$ as a function of $\beta$
for the asymmetric cross; 
the solid line is the least squares fit 
$\left. E_1^{({\rm eo})}/E_{\rm TH} \right|^{\rm FIT} = 3.96521 +\frac{11.0968}{\beta ^{7.38923}}-\frac{10.156}{\beta ^{3.69462}}$.}
\bigskip
\label{fig_eo1}
\end{center}
\end{figure}

The  ratio $E^{(\rm eo)}/E_{\rm TH}$  is plotted in Fig.\ref{fig_eo1} 
and compared with the fit 
\[\left. E_1^{({\rm eo})}/E_{\rm TH} \right|^{\rm FIT} = 3.96521 +\frac{11.0968}{\beta ^{7.38923}}-\frac{10.156}{\beta ^{3.69462}}.\]
As for the even-even state, this behavior is consistent with the survival of the bound state for $\beta \rightarrow \infty$.

\begin{figure}[ht]
\bigskip
\begin{center}
\includegraphics[scale=0.27]{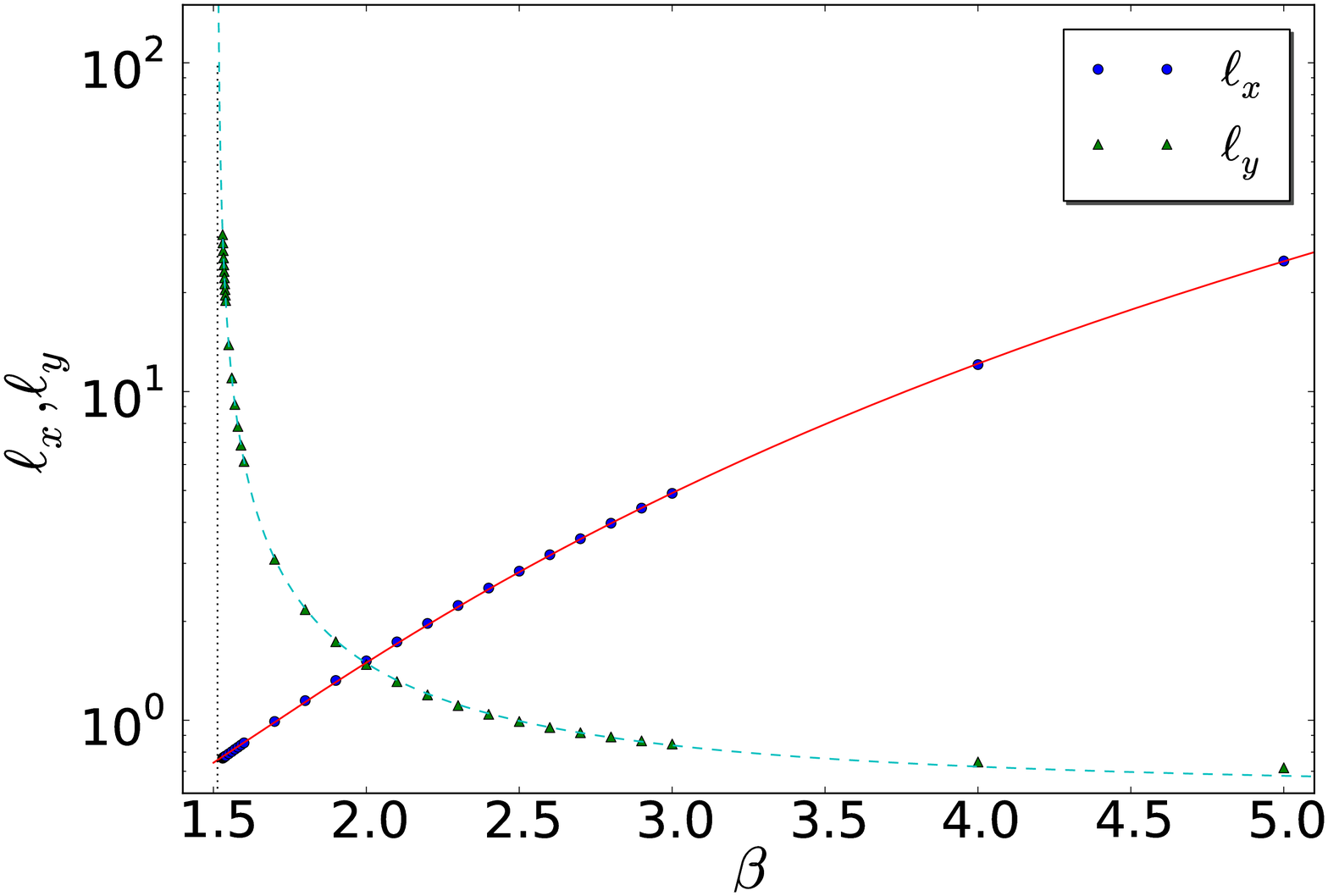}
\caption{(color online) $\ell_x$ and $\ell_y$ as a function of $\beta$
for the lowest even-odd state of the asymmetric cross; 
the solid and dashed lines are the least squares fits
$\left. \ell_x \right|^{\rm FIT} = 0.129087 \beta ^{3.26312}+0.258401$ and
$\left. \ell_y \right|^{\rm FIT} = {0.604788}/({1 -{2.16467}{\beta ^{-1.86384}}})$. The vertical line corresponds to
the critical value $\beta^{({\rm eo})}_\star = 1.513$ where $\left. \ell_y \right|^{\rm FIT}$ is singular.}
\bigskip
\label{fig_eo2}
\end{center}
\end{figure}

In Fig. \ref{fig_eo2} 
we plot $\ell_x$ and $\ell_y$ and compare with them the fits
\[\left. \ell_x \right|^{\rm FIT} = 0.129087 \beta ^{3.26312}+0.258401\] and
\[\left. \ell_y \right|^{\rm FIT} = \frac{0.604788}{1 -\frac{2.16467}{\beta ^{1.86384}}}.\]
The singularity of $\left. \ell_y \right|^{\rm FIT}$ is located at $\beta^{({\rm eo})}_\star = 1.513$ 
(the vertical line in the plot) and represents the critical value where the bound state appears.
Notice also that $\left. \ell_x \right|^{\rm FIT} \approx \left. \ell_y \right|^{\rm FIT}$ for $\beta = 2$.

\subsection{Odd-even state}

In the case of the odd-even states we have not been able to find a bound state for any of the values of $\beta$ considered. 
Once again this behavior is consistent with our earlier predictions.

\section{Conclusions}
\label{conclusions}

For years now, it has been verified theoretically, as well as experimentally, 
that the spectra of waves or quantum particles moving in several open two-dimensional systems could exhibit bound states. 
Since these bound states arise from the symmetry in the system,
for applications it is important to find out how the perturbation of the given symmetry affects the corresponding spectrum.

In this paper we have shown that the question about the role of the symmetry of several setups can be answered by studying the simple problem of an effective one-dimensional quantum well. 
We support this conclusion by means of very accurate numerical solutions of the full Helmholtz equation with Dirichlet boundary conditions.
It is worth to notice that the conclusions drawn from the simple problem are valid even for values of the ratio of the width of the crossing branches ($\beta$)
for which the one-dimensional approximation not longer holds.

In particular, we have shown that the bound state of the symmetric cross persists for any finite value of $\beta$.
A similar, but richer, behavior is obtained for the T-shaped configuration as the width of the vertical bar of the T tends to zero (i.e., $\beta \to \infty$):
from $\beta=1$ up to $\beta_\star=1.513$ the wave function is unbound; 
above that critical value the wave function becomes bounded, but mainly localized in the vertical bar of the T;
finally, if $\beta$ is still increased, the wave function remains bound, but mainly localized in the horizontal bar of the T.
Of course, in any of these two setups, the bound state must dissapear as soon as the width of the decreasing branch reaches zero.
Therefore, this seems to imply that adding just a very small perturbation on a very large one-dimensional quantum wire or waveguide 
may have a striking impact on the transport of charge carriers or waves along it.

The situation is different for the other two cases we analyzed.
For the L-shaped configuration we have found that the bound state becomes unbounded at the critical value $\beta_\star=1.2279$ of the ratio of the width of the intersecting branches.
Finally, for the T-shaped configuration no bound state arises when the width of the horizontal arm is decreased towards zero.

How strong are these effects as to be observable and appliable will depend on how strong is the departure of the actual experimental setup from the ideal configurations considered. 
Nevertheless, if the dependence on the symmetry found here is confirmed by future experiments, then it could be used as a switch-like mechanism for trapping and untrapping waves and charge carriers.
This would give rise to a number of important applications.

\section{Acknowledgments}
This research was supported by the Sistema Nacional de Investigadores (M\'exico). 
The work of CAT-E was also partially funded by PROMEP (M\'exico) under grant PROMEP/103.5/10/4948.

\end{document}